\documentstyle[12pt,rotate]{article}

\input epsf.sty

\newcommand{\inclfig}[2]{\mbox{\epsfxsize=#1cm \epsfbox{#2.eps}}}
\newcommand{\F}{{\cal F}}

\newcommand{\GeVsq}{{\rm GeV}^2}

\newcommand{\k}{{\kappa}}

\begin{document}
\begin{titlepage}

{\hfill \parbox{20mm}{{ NTZ 97/31\\[2mm]
                        TPR-97-22}}}

\vspace{20mm}

\centerline{\large \bf On the leading logarithmic evolution of the
                       off-forward distributions}

\vspace{18mm}

\centerline{\bf A.V. Belitsky$^{1,2}$, B. Geyer$^3$, D. M\"uller$^3$,
                A. Sch\"afer$^2$}

\vspace{18mm}

\centerline{\it ${^1}$Bogoliubov Laboratory of Theoretical Physics,
                Joint Institute for Nuclear Research}
\centerline{\it 141980, Dubna, Russia}
\centerline{\it ${^2}$Institut f\"ur Theoretische Physik, Universit\"at
                Regensburg}
\centerline{\it D-93040 Regensburg, Germany}
\centerline{\it ${^3}$Institute for Theoretical Physics, Center of
                Theoretical Science, Leipzig University}
\centerline{\it  04109 Leipzig, Germany}

\vspace{25mm}

\centerline{\bf Abstract}

\hspace{0.8cm}
{\ }

{\noindent
We have found the analytical solution of the LO-evolution equation
for off-forward distributions which arise in the processes of deeply
virtual Compton scattering or exclusive production of mesons. We
present the predictions for their evolution with an input distribution
taken from recent bag model calculations.
}

\vspace{0.5cm}

\end{titlepage}

\begin{figure}[htb]
\begin{center}
\vspace{3.8cm}
\hspace{-2.5cm}
\mbox{
\begin{picture}(120,20)(100,100)
\put(0,-30)                    {
\epsffile{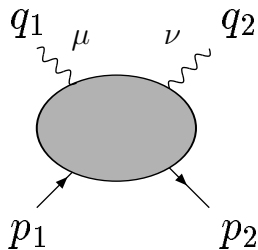}
                               }
\put(180,218){$\mu$}
\put(215,218){$\nu$}
\end{picture}
}
\end{center}
\vspace{-1.5cm}
\caption{ \label{DVCSkinematics} Generic form of the DVCS amplitude.}
\end{figure}

\noindent {\it 1. Introduction.} Recently, X. Ji proposed to
explore deeply virtual Compton scattering (DVCS) (see Fig.\
\ref{DVCSkinematics}) to get deeper insight into the spin
structure of the polarized nucleon and, namely, to determine
the total quark angular momentum (spin and orbital angular
momentum) in the nucleon \cite{Ji97,Ji97a}. It turned out that
this process is very interesting in its own right, since it
allows to obtain information about some non-perturbative
off-forward parton distributions (OFPDs), which are inaccessible
in ordinary inclusive measurements such as deep inelastic
scattering (DIS). Planning of experiments to measure DVCS is
under way, but it is hampered by the lack of reliable theoretical
predictions in the interesting ($\omega$, $\eta$, $Q^2$)-range.
A crucial step towards this goal is to understand the
$Q^2$-evolution of OFPD. In the present short note we, therefore,
present the $Q^2$-evolution of the nonpolarized quark distribution
function in the leading order (LO) approximation starting form MIT
bag model predictions at a low energy scale \cite{JiMelSon97}.

To make this paper self-contained, as much as possible, let us first
recall the main points of the perturbative (LO) analysis
\cite{MRGDH94,Ji97,Rad96,CFS96,BGR97,FFGS97,Che97,BM97a} of DVCS
amplitude. In the configuration space such a Compton process, given
by the following expression
\begin{eqnarray}
T^{\mu\nu}(\omega,\eta,Q^2)
= i \int d^4z\, e^{izQ}
\langle p_2 |\,
T\left\{ J^\mu\left( - z/2 \right) J^\nu \left( z/2 \right) \right\}
| p_1 \rangle ,
\end{eqnarray}
is dominated by singularities of the $T$-product of electromagnetic
currents $J_\mu$ on the light-cone. In momentum space this dominance
corresponds to the generalized Bjorken limit:
\begin{eqnarray}
- Q^2 \to \infty,
\qquad PQ \to \infty,
\qquad \omega,\Delta^2 \mbox{\ -fixed},
\nonumber
\end{eqnarray}
where the variables are defined as follows $Q= \frac{1}{2}(q_1+q_2)$,
$P=(p_1 + p_2)$, $\Delta = p_1 - p_2 = q_2 - q_1$. DVCS is
characterized by two scaling variables $\omega = - \frac{PQ}{Q^2}$ and
$\eta = \frac{\Delta Q}{PQ}$ \cite{MRGDH94}, the former is analogous
to the Bjorken one in ordinary DIS while the latter gives the
magnitude of the skewedness of the process in question and
parametrizes different high-energy two-photon reactions according
to the Table \ref{tab-KinLCd}.
\begin{table}[hbt]
\begin{center}
\begin{tabular}{|l|c|l|}
\hline
DIS              & $\Delta = 0$  & $\quad\eta  = 0$           \\
DVCS             & $q_2^2=0$     & $\quad \eta = 1/\omega$    \\
$\gamma^\ast N
\to Nl^+ l^-$    & $q_2^2>0$     & $\quad\eta\approx
                                   \cos\phi_{\rm Br}
                                   = {{\bf p}_1{\bf Q}
                                   /|{\bf p}_1||{\bf Q}|}$,
                                   Breit--frame               \\
$\gamma^\ast
\gamma^\ast
\to H H$         &               & $\quad\eta\approx
                                   \cos\phi_{\rm cm}
				   = {{\bf p}_1{\bf Q}
				   /|{\bf p}_1||{\bf Q}|}$,
				   center-of-mass frame       \\
$\gamma^\ast
\gamma^\ast
\to M$           & $p_1=0$       & $\quad\eta=1$              \\
\hline
\end{tabular}
\end{center}
\caption{\label{tab-KinLCd} The value of the scaling variable $\eta$
for different two-photon processes.}
\end{table}

Notice that in general the unpolarized tensor $T^{\mu\nu}$ can be
decomposed into four (three in the case of DVCS) independent
kinematical structures that separately respect current conservation
(similarly for polarized hadrons) \cite{Lu97}. However, only two of
them survive the general Bjorken limit, while the remaining ones
are kinematically suppressed and contain higher twist contributions.
In LO only one kinematical structure is independent, so that the
longitudinal helicity amplitudes vanish. This is analogous to the
Callan-Gross relation in unpolarized DIS. Motivated by the result
that only transverse helicity amplitudes contribute in LO, we can
write, e.g. for unpolarized scattering
\begin{eqnarray}
\label{DVCS-ampl-unpol}
T(\lambda',\lambda)
&=& \epsilon_2^\mu(\lambda')
T_{\mu\nu}
\epsilon_1^\nu(\lambda)
\approx
\frac{1}{2} \epsilon_2^\nu (\lambda ') \epsilon_1^\nu (\lambda)\,
\delta_{\lambda 1/2}\, \delta_{\lambda' 1/2}\,
T_{\mu\mu}.
\end{eqnarray}
Adding the surviving spin-dependent structure, we get\footnote{Recently,
all NLO coefficient functions \cite{JiOs97,BM97a} as well as
$\alpha_s$-corrections to the eigenfunctions of the two-loop
$QQ$- and $QG$-evolution kernels became available.}:
\begin{eqnarray}
T^{LO}_{\mu\mu}
&=& \int_{-1}^1 dt
\sum_i Q_i^2 q^i \left( t, \eta, \Delta^2 \right)
\left\{
\frac{\omega}{\omega t - 1 + i0} + \frac{\omega}{\omega t + 1 - i0}
\right\}, \nonumber\\
\label{DVCS-ampl-pol}
T^{LO}_{[\mu\nu]}
&=& \frac{ i \epsilon_{\mu\nu P Q}}{2PQ}
\int_{-1}^1 dt
\sum_i Q_i^2 \Delta q^i\left( t, \eta, \Delta^2 \right)
\left\{
\frac{\omega}{\omega t - 1 + i0}
- \frac{\omega}{\omega t + 1 - i0}
\right\}.
\end{eqnarray}
Here the off-forward distributions are defined in terms of the
light-cone Fourier transformation of the renormalized twist-2
light-ray operators:
\begin{eqnarray}
\label{def-non-for}
\left\{
\begin{array}{c}
q \\
\Delta q
\end{array}
\right\}
\left( t, \eta, \Delta^2, \mu^2 \right)
= \int \frac{d\k}{2\pi} e^{\frac{i}{2} \k t P_+ }
\langle p_2 |
\bar{\psi} (- \k /2 \, n)
\left\{
\begin{array}{c}
\gamma_+  \\
\gamma_+ \gamma_5
\end{array}
\right\}
\left.
\psi (\k /2 \, n) \, \right|_{\mu^2}
| p_1 \rangle_{\Delta_+ = \eta P_+} ,
\end{eqnarray}
and we have omitted the path ordered link factor which ensures gauge
invariance. The integration range of the variable $t$ in
Eq.\ (\ref{DVCS-ampl-unpol}) results from the spectral properties of
the off-forward distribution amplitude that can be derived in a
straightforward manner with the help of Jaffe's approach \cite{Jaffe83}
for studying the support properties of parton densities. In our
conventions the variable $t$ has no direct interpretation as
longitudinal momentum fraction, but it has the advantage that its
region is process independent: $-1 \leq t \leq 1$. The same feature
is appropriate to the quark $\F_\zeta^a ( x ) $ ($0 \leq x \leq 1$)
and anti-quark $\F_\zeta^{\bar a}(x)$ ($-1 \leq x \leq 0$) non-forward
distributions introduced by Radyushkin \cite{Rad96}. The
$\zeta$-dependent limits appear only when one attempts to combine
them into one function $\widetilde \F_\zeta^a(x) = \theta (x)
\theta (\bar x) \F_\zeta^a(x) - \theta (\zeta - x) \theta (x + 1 - \zeta)
\F_\zeta^{\bar a} (\zeta - x)$ which is exactly the OFPD introduced
above in Eq. (\ref{def-non-for}) in different notations. The
non-diagonal distribution\footnote{In the Radyushkin's conventions
$f^g (x, x - \zeta) = \widetilde \F_\zeta^g (x)/ x (x - \zeta)$
\cite{Rad96} .} $f^g (x_1, x_2)$ of Collins, Frankfurt and Strikman
\cite{CFS96} is parametrized in terms of the momentum fraction of
the considered partons and consequently their limits depend on the
skewedness of the processes. Thus, the momentum fractions of the
incoming and outgoing partons are $x = \frac{t + \eta}{1 + \eta},$
and $x - \zeta = \frac{t - \eta}{1 + \eta}$, respectively, with
$\zeta = \frac{2\eta}{1 + \eta}$. For $t > \eta$ both momentum
fractions are positive and, therefore, a quark field with longitudinal
momentum $x p_1$ arise from the incoming hadron and enters with
momentum $( x - \zeta ) p_1$ into the outgoing hadron. For $- \eta > t$
both momentum fractions are negative and the analogous picture for
two anti-quarks holds true. In these regions, the corresponding
off-forward quark and anti-quark distributions merge in the forward
case to the well-known quark and anti-quark distributions with
support $0 \le t \le 1$ and $-1 \le t \le 0$, respectively.
For $-\eta < t < \eta$, $x$ is positive while $x - \zeta$ is
negative, so that both partons come from the same hadron. Here the
OFPD looks like a distribution amplitude. Indeed, inserting the vacuum
and one-meson state in the definition (\ref{def-non-for}) instead of
incoming and outgoing nucleon state, we recover the definition of the
meson distribution amplitude depending on the momentum fraction
$0\le x=(1+t)/2 \le 1$, where $\eta = 1$ \cite{ER-BL}. Thus, OFPDs
are hybrids and their probability interpretation depends upon the
interplay between the respective values for $t$ and $\eta$.

Finally, to end up with this introduction, we point out a few
difficulties of the measurement of the OFPDs in DVCS. DVCS can be
explored in the reaction
\begin{eqnarray}
e^\pm N \to e^\pm \gamma N.
\end{eqnarray}
However, DVCS and Bethe-Heitler (BH) sub-processes have the same
final states. Hence, they interfere and the measured cross section
contains the squares of pure DVCS and BH parts as well as the
interference term:
\begin{eqnarray}
d\sigma \propto |T_{\rm DVCS}|^2
+ |T_{\rm BH}|^2
+ \left(T^*_{\rm DVCS} T_{\rm BH} + T^*_{\rm BH} T_{\rm DVCS}\right).
\end{eqnarray}
Fortunately, the contribution from the BH process is known from
the measurements of the nucleon form factors. In unpolarized or double
spin experiments DVCS amplitude is given as convolution [compare with
Eqs.\ (\ref{DVCS-ampl-unpol}) and (\ref{DVCS-ampl-pol})], so that a
direct extraction of the OFPDs seems to be very difficult\footnote{
The imaginary part of the LO scattering amplitude (\ref{DVCS-ampl-pol})
is expressed in terms of the OFPDs at $t = \pm 1/\omega$, while the 
real part is given as a Hilbert transformation and can be inverted in
principle.}. In such experiments the BH contribution dominates over
DVCS for small $\Delta^2$. Its importance decrease, however, with
increasing $\Delta^2$ \cite{Ji97}. In single spin experiments a
non-vanishing cross section arises from the imaginary part of the
amplitude. Since BH has only a real part, the single spin asymmetry
is proportional to the imaginary part of the DVCS amplitude, so that
the OFPDs can be measured directly at the point $t = \pm 1/\omega$.
Hopefully, as a first step in the direction of DVCS measurements,
the predicted scaling behaviour in the generalized Bjorken regime
will be investigated at CEBAF \cite{Ji97,JiMelSon97}.


\noindent {\it 2. Evolution equation.} The evolution of OFPDs arises
technically from the renormalization procedure for the light-ray
operators \cite{AniZav78}, which satisfy in the non-singlet channel
the well-known renormalization group equation
\cite{Geyer82,Balitsky83,BraGeyRob87,BalBra89}. Employing the
definition (\ref{def-non-for}), it is straightforward to derive
from the latter the evolution equation for the OFPD:
\begin{eqnarray}
\label{EvoEqu-nonfor}
Q^2\frac{d}{dQ^2}
q^{\rm NS}\left( t,\eta,\Delta^2,Q^2 \right)
= \frac{\alpha_s}{2 \pi}
\int_{-1}^{1} \frac{dt'}{|\eta|}
\left[ V \left( \frac{t}{\eta}, \frac{t'}{\eta} \right) \right]_+
q^{\rm NS}\left(t',\eta,\Delta^2,Q^2\right).
\end{eqnarray}
In the restricted region $|t,t'|\le 1$ the kernel $V (t,t')$
coincides with the Efremov-Radyushkin-Brodsky-Lepage (ER-BL) one
\cite{ER-BL}:
\begin{eqnarray}
V_{\rm ER-BL} ( x , y )
= \left. V( 2 x - 1 , 2 y - 1 ) \right|_{0 \le x , y \le 1} .
\end{eqnarray}
Moreover, the continuation to the whole $t,t'$-plane is unique
\cite{MRGDH94}, so that the extended evolution kernel $V (t , t')$
can be easily recovered from the corresponding ER-BL analogues. From
the support restrictions and charge conjugation symmetry a
representation was derived in \cite{DitMueRobGeyHor88}, which is
valid in leading order:
\begin{equation}
\label{GAMTRE}
V (t, t') = \Theta^0_{11} ( t - t', t - 1) {\cal V} (t, t')
+ (t \to - t, t' \to - t'),
\end{equation}
where $\Theta^0_{11} (x_1 , x_2) = \bigl( \theta(x_1)\theta(- x_2) -
\theta(x_2)\theta(- x_1) \bigr) / ( x_1 - x_2 )$, and ${\cal V} (t, t')$
is an analytic function of $t$ and $t'$. Thus, the extended kernel can
be obtained from the exclusive ER-BL one by replacing the ordinary
$\theta$-functions in front of ${\cal V} (t, t')$ by the generalized
ones, namely, $\theta (t - t') / (1 - t') \to \Theta^0_{11}
( t - t', t - 1)$. For concrete examples the reader is referred to
Ref. \cite{BM97a}. Beyond LO, further contributions appear in Eq.\
(\ref{GAMTRE}), but they are not of relevance for our present
consideration.

For $\eta\to 0$ the evolution equation (\ref{EvoEqu-nonfor}) turns
into the DGLAP one with a splitting function given by the limit:
\begin{eqnarray}
P(z)=\lim_{\eta\to 0} \frac{1}{|\eta|}
\left[
V \left( \frac{z}{\eta}, \frac{1}{\eta} \right)
\right]_+ .
\end{eqnarray}


\noindent {\it 3. Solution of the evolution equation.} The evolution
equation analogous to Eq.\ (\ref{EvoEqu-nonfor}), but for gluons, was
attempted to be integrated numerically \cite{FFGS97}. However, it
remains an important issue to find an analytical solution. The way to
do that is based on the use of the conformal invariance of the free
field theory which ensures that conformal 2-particle operators do not
mix under renormalization at LO \cite{ER-BL,Mak81,Ohr82,Mue94,BM97a}.
Such conformal operators are given in terms of Gegenbauer polynomials
${^Q\!{\cal O}}_{jl} = \bar{\psi} (i \partial_+)^l \gamma_+
C^{\frac{3}{2}}_j \left( {\stackrel{\leftrightarrow}{D}_+}
/ {\partial_+} \right) \psi $ and their off-forward matrix elements
can be obtained from the moments
\begin{eqnarray}
\label{con-mellin}
q_j \left( \eta, \Delta^2, Q^2 \right)
= \eta^j \int_{-1}^{1} dt\, C_j^{\frac{3}{2}}
\left( \frac{t}{\eta} \right) q \left( t, \eta, \Delta^2, Q^2 \right)
= \frac{2}{P_+^{j + 1}}
\left. \langle p_2 |
{^Q\!{\cal O}}_{jj} (0) \right|_{\mu^2 = Q^2}
| p_1 \rangle .
\end{eqnarray}
In the following, we restrict ourselves to the flavour non-singlet
channel since up to now no estimation of the off-forward gluon
distribution has been performed yet as long as available strong
interaction models do not contain the gluon fields at all. The
generalization of our results to the singlet channel is
straightforward. The only complication which is due to quark-gluon
mixing can be treated in the same way as in DIS. For $\eta = 1$
it is well-known that the Gegenbauer polynomials diagonalize the
ER-BL-kernel. Moreover, the support of the extended evolution
kernel ensures this property for arbitrary $\eta$:
\begin{eqnarray}
\int_{-1}^1 dt\,
C_j^{\frac{3}{2}}
\left( \frac{t}{\eta} \right)
\left[
V \left( \frac{t}{\eta}, \frac{t'}{\eta} \right)
\right]_+
=\gamma_j
C_j^{\frac{3}{2}}
\left( \frac{t'}{\eta} \right) .
\end{eqnarray}
The remaining problem is to find the inverse Mellin transformation of
the moments (\ref{con-mellin}). The support property\footnote{In the
subsequent discussion we omit, for brevity, the dependence of the
distributions on the momentum transferred squared.} of $q^{\rm NS}
\left( t, \eta, Q^2 \right)$ allows to expand it with respect to
an appropriate complete set of polynomials $C^\nu_k(t)$ which are
orthogonal in the domain $-1\le t\le 1$ with the weight function
$w(t | \nu) = (1 - t^2)^{\nu - \frac{1}{2}}$:
\begin{eqnarray}
\label{NonForEvo}
q^{\rm NS} \left( t, \eta, Q^2 \right)
= \sum_{j=0}^{\infty} \frac{w(t | \frac{3}{2})}{N_j(\frac{3}{2})}
C^\frac{3}{2}_j (t) a_j ( \eta, Q^2 | {\scriptstyle \frac{3}{2}} ),
\end{eqnarray}
where $N_j(\nu)= 2^{ - 2 \nu + 1 }
\frac{ \Gamma^2 (\frac{1}{2}) \Gamma ( 2 \nu + j )}
{\Gamma^2 (\nu) ( \nu + j ) j! }$ is a normalization factor. It is
straightforward to calculate the expansion coefficients in terms
of the conformal moments (\ref{con-mellin}):
\begin{eqnarray}
\label{a-j}
a_j ( \eta, Q^2 | {\scriptstyle \frac{3}{2}} ) =
\sum_{k=0}^j a_{jk} ( \eta | {\scriptstyle \frac{3}{2}} )
q^{\rm NS}_k \left( \eta, Q^2_0 \right)
\exp
\left\{
\gamma_k
\int_{Q_0^2}^{Q^2} \frac{d\sigma}{\sigma}
\frac{\alpha_s(\sigma)}{2\pi}
\right\},
\end{eqnarray}
where $\gamma_j = C_F [ - 2 \psi(j + 2) + 2 \psi(1) + 1/(j+1) - 1/(j+2)
+ 3/2 ]$ are the well-known forward non-singlet anomalous dimensions,
and $a_{jk}(\eta | {\scriptstyle \frac{3}{2}} )$ can be written in a
very compact manner by employing the definition of hypergeometric
functions\footnote{To be general we have not specified the index
$\nu$: for the quark operators $\nu=3/2$, for gluons $\nu=5/2$.}:
\begin{eqnarray}
\label{a-jk}
a_{jk}( \eta | \nu )
&=&
\int_{-1}^{1} dt \frac{w (t | \nu)}{ \eta^k N_k (\nu) }
C^\nu_k ( t ) C^\nu_j ( \eta t )
\nonumber\\
&=&
\frac{1}{2} \theta_{jk} \left[ 1 + (-1)^{j - k} \right]
\frac{ (-1)^{\frac{j - k}{2}}
\Gamma \left( \nu + \frac{j + k}{2} \right) }
{ \Gamma \left( \nu + k \right)
\Gamma \left( 1 + \frac{j - k}{2} \right) }\,
{_2F_1}
\left( \left. {\frac{k - j}{2}, \nu + \frac{j + k}{2} \atop \nu + k + 1}
\right| \eta^2 \right).
\end{eqnarray}

Let us turn to the consideration of the limiting cases. Since
${_2F_1}( - n, a + n; a + 1; 1 ) = \delta_{ n 0 }$, we obtain for
$\eta = 1$:
\begin{eqnarray}
a_k( 1, Q^2 | {\scriptstyle \frac{3}{2}})=
q^{\rm NS}_k\left( 1, Q^2_0 \right)
\exp
\left\{
\gamma_k
\int_{Q_0^2}^{Q^2}
\frac{d\sigma}{\sigma}
\frac{\alpha_s(\sigma)}{2\pi}
\right\},
\end{eqnarray}
and, therefore, from Eq.\ (\ref{NonForEvo}) we recover the well-known
LO result for the evolution of the meson distribution amplitude
\cite{ER-BL}. In the limit $\eta\to 0$, the solution of the DGLAP
equation expanded with respect to Gegenbauer polynomials
$C_k^\frac{3}{2}$, has the following coefficients
\begin{eqnarray}
a_j (0, Q^2 | {\scriptstyle \frac{3}{2}} )
= \sum_{{k=0 \atop k - j {\rm \ even}}}^j
\frac{ (-1)^{\frac{j - k}{2}} \Gamma\left( \frac{3 + j + k}{2} \right)}
{\Gamma\left( \frac{3 + 2 k}{2} \right)
\Gamma\left(\frac{2 + j - k}{2}\right)}
q^{\rm NS}_k \left( 0, Q^2_0 \right)
\exp
\left\{
\gamma_k
\int_{Q_0^2}^{Q^2}
\frac{d\sigma}{\sigma}
\frac{\alpha_s(\sigma)}{4\pi}
\right\}.
\end{eqnarray}
In this limit conformal moments coincide up to an overall normalization
with the usual ones in DIS:
\begin{eqnarray}
q^{\rm NS}_k \left( 0, Q^2_0 \right)=
\frac{ 2^k \Gamma( \nu + k ) }{ k! \Gamma(\nu) }
\int_{0}^1 dx x^k \left[q^{\rm NS}\left(x,0,Q^2_0\right)+
(-1)^k q^{\rm NS}\left(-x,0,Q^2_0\right) \right].
\end{eqnarray}

\begin{figure}[htb]
\unitlength1mm
\begin{center}
\vspace{-2.0cm}
\hspace{0.0cm}
\mbox{
\begin{picture}(90,60)
\put(0,0){\inclfig{7}{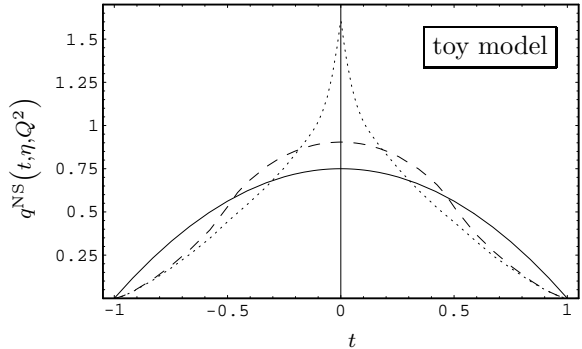}}
\put(-6,13){\rotate{$\scriptstyle q^{{\rm NS}}\left(t,\eta,Q^2\right)$}}
\put(39,-4){$\scriptstyle t $}
\put(49,35){\fbox{\footnotesize toy model}}
\end{picture}
}
\end{center}
{\caption{ \label{Fig-EvNon} The toy input OFPD (solid line) given at
$Q^2=0.5\ \GeVsq$ is evolved up to  $Q^2=5\ \GeVsq$, where
$\Lambda_{\rm QCD} = 220$ MeV. The dashed and dotted lines show
the evolution for $\eta = 0.5$ and $\eta=0$, respectively, while for
$\eta = 1$ the input distribution does not evolve since it has already
the asymptotic shape.}
}
\end{figure}

For asymptotically large $Q^2$ all conformal moments with $k > 0$ will
be suppressed due to non-zero anomalous dimensions and, consequently,
only $j$-even expansion coefficients survive in this limit:
\begin{eqnarray}
a_j^{\rm as}( \eta | {\scriptstyle \frac{3}{2}} )
&=& \lim_{Q\to\infty} a_j ( \eta, Q | {\scriptstyle \frac{3}{2}} ) \\
&=&
\frac{(-1)^{\frac{j}{2}}\Gamma\left(\frac{3 + j}{2}\right)}
{\Gamma\left(\frac{3}{2}\right)\Gamma\left(\frac{2 + j}{2}\right)}
{_2F_1}
\left(
\left.
{ - j/2, ( 3 + j )/2 \atop 5/2 }
\right|
\eta^2
\right)
q_0 \left( \eta, Q^2_0 \right) \quad \mbox{for $j$ even}.
\nonumber
\end{eqnarray}
It is not hard to check that they belong to the following asymptotic
off-forward distribution which was originally found in Ref.
\cite{Rad96}:
\begin{eqnarray}
q^{{\rm as}}\left(t,\eta\right)
\propto
\frac{1}{|\eta|}
\theta \left( 1 - \frac{t^2}{\eta^2} \right)
\left( 1 - \frac{t^2}{\eta^2} \right).
\end{eqnarray}
It turns into $\delta(t)$ when $\eta \to 0$. From the asymptotic
form found so far, we can argue that, in general, OFPDs will be enhanced
in the region $0 \le |t| < \eta$ and suppressed for $\eta < |t| \le 1$
when evolved upwards in $Q^2$. To demonstrate these effects in a clear
manner, we choose a $\eta$-independent toy-model input distribution
$3/4 (1-t^2)$ (which is in fact unphysical for arbitrary $\eta$).
For $\eta = 1$ this distribution possesses already the asymptotic form,
while for smaller $\eta$'s the evolution provides the mentioned behaviour
as illustrated in Fig.\ \ref{Fig-EvNon}.
The evolution has been done with the help of
Eqs.\ (\ref{NonForEvo})--(\ref{a-jk}), while the conformal
moments of our toy distribution at the input scale $Q^2=0.5\ \GeVsq$
was computed from Eq.\ (\ref{con-mellin}). Taking the first 80 terms
in Eq.\ (\ref{NonForEvo}) the OFPD was evolved to $Q^2=5\ \GeVsq$ for
$\eta=\{5/10,0\}$ and $t=0,\pm 1/20, \pm 2/20,\dots,1$. To avoid
numerical problems arising from the oscillation of the Gegenbauer
polynomials the calculation were done exactly; and, finally, the
result was interpolated to a smooth function.


\noindent {\it 4. $Q^2$-evolution of the bag model motivated OFPD.} A
first non-perturbative estimation of the off-forward valence quark
distribution function has been done using the MIT bag
model\footnote{Quite recently there appeared a paper \cite{Pol97}
with evaluation of the OFPDs in the instanton inspired chiral
quark-soliton model of the nucleon. However, it is not possible to
use their results here as only flavour singlet quark density was
considered. Moreover, flavour non-singlet combination is
$1/N_c$-suppressed in this approach.} \cite{JiMelSon97}. In this
model calculation it turned out that at a scale $\mu^2_{\rm bag}
\simeq 0.2\ \GeVsq$ the $\eta$-dependence of the off-forward $u$ and
$d$ quark distributions is extremely weak\footnote{This is in
contrast with the result of Ref. \cite{Pol97}, where the
$\eta$-dependence is rather strong.} and these functions vanish for
negative $t$ as well as for $t\to 1$. (The nonzero result of the
calculated $t$-dependence in the vicinity of unity is a model artifact,
which reflects the fact that the incoming and outgoing protons are not
good momentum eigenstates.) For positive $t$ the distributions are
positive with a maximum at $t\approx 0.4$. To be able to satisfy the
sum rules which are obeyed by the off-forward distributions, namely,
the first two moments should be independent of the skewedness
parameter $\eta$
\begin{eqnarray}
\int_{-1}^1 dt\,
\left\{
\begin{array}{c}
1 \\
t
\end{array}
\right\}
q (t, \eta, Q^2)
= \frac{2}{P_+^2}
\langle p_2 |
\left\{
\begin{array}{c}
P_+\, J^+ (0) \\
{^Q {\mit \Theta}^{++}} (0)
\end{array}
\right\}
| p_1 \rangle ,
\end{eqnarray}
(${^Q {\mit \Theta}^{\mu\nu}}$ is a quark part of the energy-momentum
tensor), we neglect the $\eta$-dependence altogether, which is not
far from the bag model results. Since the main goal of the present study
is to acquire some intuition about the evolution properties of OFPD,
we do not pursue the aim for construction of more realistic models;
rather motivated by the given results, we take a very simple
parametrization of the OFPD in the non-singlet channel, namely
\begin{eqnarray}
\label{OFPDBagMod}
q^{\rm NS}( t, \eta, Q_0^2 = \mu^2_{\rm bag} )
= 60 \, \theta( t ) t^2 (1 - t)^3,
\end{eqnarray}
with a first moment normalized to unity.

\begin{figure}[htb]
\unitlength1mm
\begin{center}
\begin{picture}(150,110)(0,0)
\put(0,60){\inclfig{6}{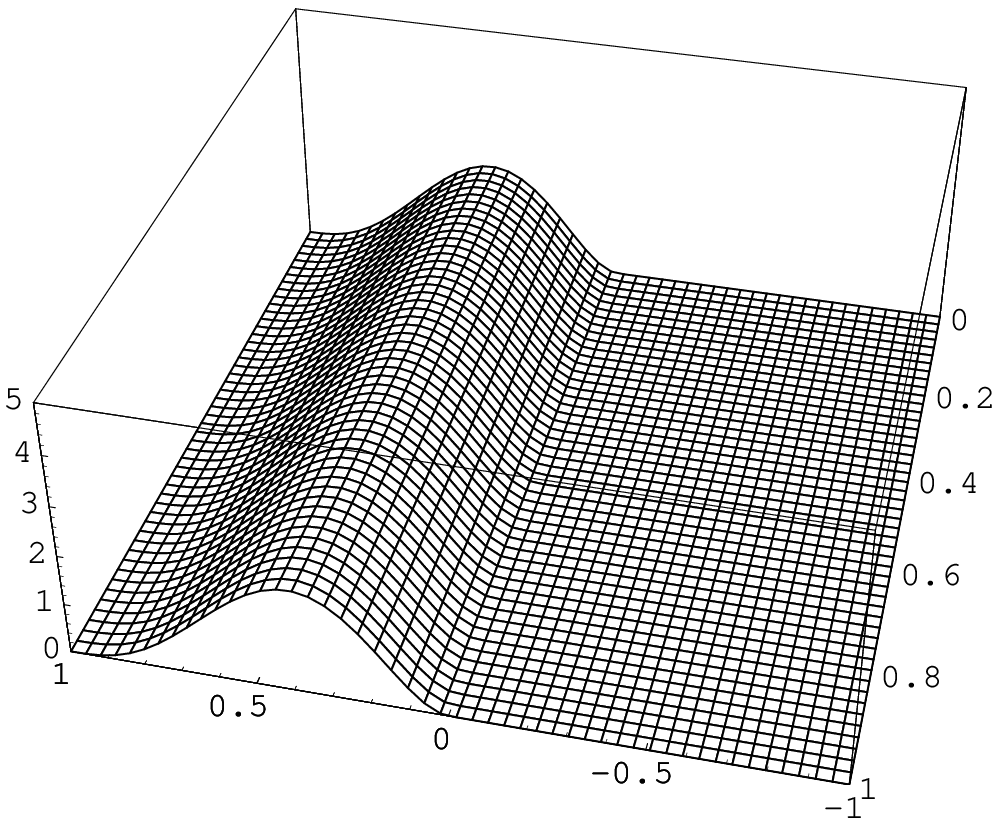}}
\put(50,95){(a)}
\put(60,75){$\eta$}
\put(-5,67){\rotate{$\scriptstyle q^{{\rm NS}}\left(t,\eta,Q^2\right)$}}
\put(80,60){\inclfig{6}{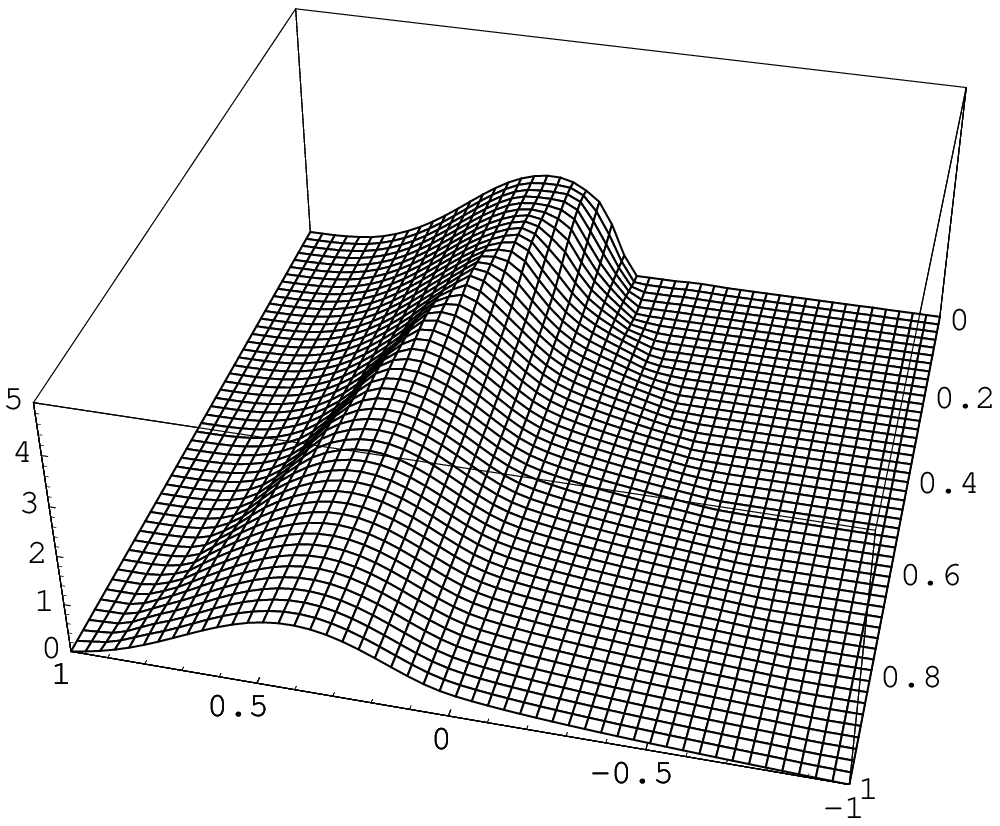}}
\put(130,95){(b)}
\put(0,0){\inclfig{6}{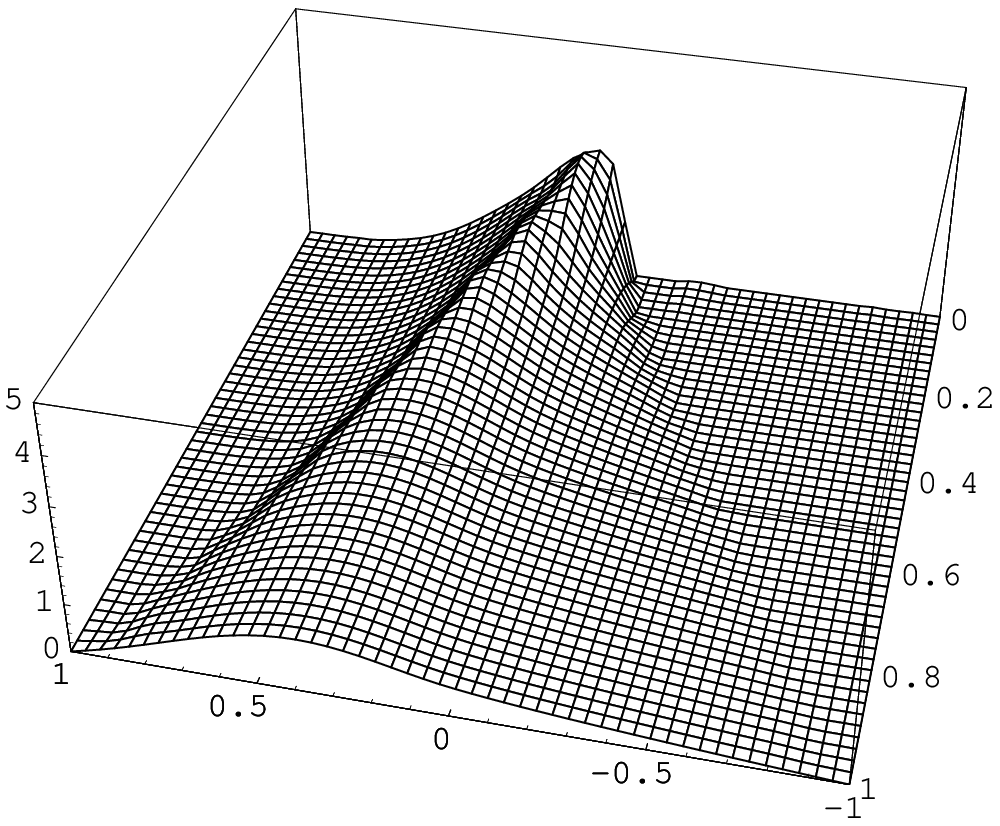}}
\put(50,35){(c)}
\put(27,0){$t$}
\put(60,15){$\eta$}
\put(-5,7){\rotate{$\scriptstyle q^{{\rm NS}}\left(t,\eta,Q^2\right)$}}
\put(80,0){\inclfig{6}{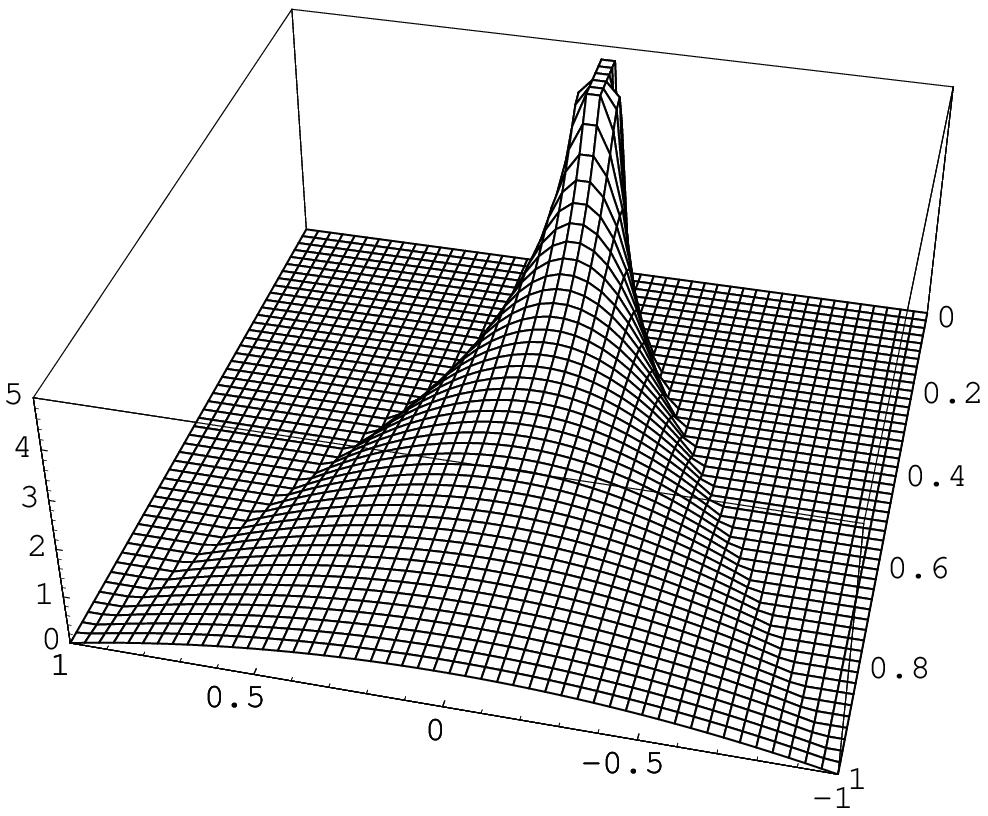}}
\put(130,35){(d)}
\put(107,0){$t$}
\end{picture}
\end{center}
\caption{ \label{Fig-EvNoBM} The evolution of the bag model
motivated OFPD (\protect\ref{OFPDBagMod}) for
$\Lambda_{\rm QCD}=220$ MeV, where the scales are $Q_0^2
= \mu^2_{\rm bag}$ (a), $Q^2 = 2\ \GeVsq$ (b), $Q^2=200\ \GeVsq$
(c), and asymptotically large $Q^2$ (d).}
\end{figure}

In Fig.\ \ref{Fig-EvNoBM}, we evolve this input as described previously
(we took also discrete values for $\eta= 0,1/10,\dots,1$ and interpolated
the result with respect to $t$ and $\eta$) up to the scales $Q^2=2\
\mbox{GeV}^2$, $Q^2=200\ \mbox{GeV}^2$, and asymptotically large $Q^2$;
for other parameters, we set $N_f = 3$, $\Lambda_{\rm QCD} = 220\
\mbox{MeV}$ in $\alpha_s (Q^2)$. As expected from our previous discussion,
the distribution spreads in $t$ for larger value of $\eta$ and shrinks
for smaller ones.
These graphs clearly show that during the evolution the large-$\eta$
part spreads over the whole range of the momentum fraction while the
small-$\eta$ distribution is pushed towards $t \to 0$ and concentrates
in the vicinity of zero. Since we have omitted in the present analysis
the gluon sector, we can expect that taking it into account we will
get, probably, an even more enhanced function in the small-$t$ (and
$\eta \to 0$) region for moderate values of $Q^2$. As is easily seen
from the above figures with growing scale the OFPD approaches the
asymptotic form. Namely, for $\eta \to 1$ it takes the
$(1 - t^2)$-shape of the asymptotic distribution amplitude, while for
small $\eta$ it becomes $\delta (t)$. For intermediate $\eta$ it
smoothly interpolates between these limits. This is a  quite general
feature that should be obeyed by any reasonable (read physical)
off-forward parton density.

\vspace{0.5cm}

We would like to thank A.V. Radyushkin for careful reading of
the manuscript and useful comments and O.V. Teryaev for
discussions. A.B. was supported by the Russian Foundation for
Fundamental Research, grant N 96-02-17631 and Deutsche Physikalische
Gesellschaft Bayern. D.M. was financially supported by the Deutsche
Forschungsgemeinschaft (DFG).

\end{document}